\documentclass[12pt,preprint]{aastex}

\shorttitle{Helioseismic test of non-homologous solar radius
changes} \shortauthors{Lefebvre et al.}

\begin{document}

\title{Helioseismic test of non-homologous solar radius changes with the 11-year activity cycle}

\author{S. Lefebvre}
\affil{SAp/DAPNIA/DSM CEA Saclay, L'Orme des Merisiers B\^at. 709,
Gif sur Yvette cedex, 91191, France} \affil{UMR 7158
CEA-CNRS-Universit\'e Paris VII, 91191 Gif sur Yvette cedex, France}
\email{sandrine.lefebvre@cea.fr}
\author{A. G. Kosovichev}
\affil{W. W. Hansen Experimental Physics Laboratory, Stanford
University, Stanford, CA 94305-4085, USA} \and
\author{J. P. Rozelot}
\affil{Observatoire de la C\^ote d'Azur, GEMINI dept., Avenue
Copernic, Grasse, 06130, France}

\begin{abstract}
Recent models of variations of the Sun's structure with the 11-year
activity cycle by \citet{Sofia05} predict strong non-homologous
changes of the radius of subsurface layers, due to subsurface
magnetic fields and field-modulated turbulence. According to their
best model the changes of the surface radius may be 1000 times
larger than those at the depth of 5~Mm. We use f-mode oscillation
frequency data from the MDI instrument of Solar and Heliospheric
Observatory (SOHO) and measurements of the solar surface radius
variations from SOHO and ground-based observatories during solar
cycle 23 (1996-2005) to put constraints on the radius changes. The
results show that the above model overestimates the change of the
radius at the surface relative to the change at 5Mm.
\end{abstract}


\section{Introduction}

For the last 30 years, the question of whether the solar radius
changes with time, and with the 11-year activity cycle, in
particular, has been the subject of many controversies. Using
different techniques, several series of measurements have been made
leading so far to inconsistent results. Even if astrolabe
measurements can be considered as extracted from the same
statistical population \citep{Badache06}, \citet{Laclare96} and
\citet{Reis03} reported a variation of the solar radius in antiphase
with the solar cycle while \citet{Noel04} showed the opposite. In
addition, drift-time measurements made photoelectrically from CCD
transits at Iza$\tilde{n}$a and Locarno did not show cycle-dependent
variations in excess of $\pm$50 mas\footnote{the mas is the
milliarcsecond and is equal to about 0.725 km on the Sun}
\citep{Wittmann00}. Using SOHO/MDI images, \citet{Kuhn04} found no
significant variations at any level above 7 mas, which is
significantly lower than any variation reported from ground-based
instruments (astrolabes) ranging from 50 to 200 mas. Using the Solar
Disk Sextant experiment aboard stratospheric balloons,
\citet{Sofia94} indicated an antiphase variation between the solar
radius and the 11-year cycle.

Helioseismic methods can also be used to study the variations with
the cycle: in this case, a seismic radius is defined from
oscillation frequencies of solar f-modes \citep{Schou97, Antia98}.
The seismic radius is an interesting probe of the subsurface layers
to a depth of about 15 Mm, as recently published by \citet{Dziem04}
and \citet{Lefebvre05} (hereafter LK05).  Using properties of
$f$-mode frequencies, LK05 showed that the variations of the solar
seismic radius were nonuniform with depth. More precisely, LK05
reported a non-monotonic change in the position of the subsurface
layers: the radius changes between approximately 0.97 and
0.99~$R_{\odot}$ were in phase with the solar cycle while the radius
of the shallower layers above 0.99~$R_{\odot}$ up to the surface
changed in antiphase. In general, the results are consistent with
previous conclusions that solar-cycle variations in the solar radius
are confined in the outermost layers of the Sun \citep{Antia04,
Dziem05}.

Using an evolution code of the Sun, \citet{Sofia05} (hereafter
SBDLT05) also found a non-homologous variation of the solar radius
with depth and a radius change at the surface in antiphase with the
solar activity cycle. Using the Yale Rotating Evolution Code (YREC;
\citep{Winnick02} and including the effects of a variable magnetic
field and field-modulated turbulence, SBDLT05 found a monotonic
variation of the solar radius, sharply increasing in amplitude by a
factor of approximately 1000 from a depth of 5 Mm to the solar
surface between 1996 and 2000.

Both, the helioseismology and model results show that the
solar-cycle variations of the solar radius with depth in the
subsurface layers are non-homologous, and that the radius change at
the surface is in antiphase with the solar activity.

However, there are also discrepancies between the model and
helioseismic inferences concerning the amplitude of these variations
and the behavior of the changes in the subsurface stratification in
the first 15 Mm. LK05 found an oscillation and a change of phase in
the first 15 Mm with variations not exceeding 30 km, whereas SBDLT05
found a monotonic variation in the first 5 Mm with a factor
approximately up to 1000 between the depth at 5 Mm and the surface.
However, LK05 showed that the helioseismic inferences are not very
sensitive to the surface changes. Therefore, to make a quantitative
test of the model prediction in addition to the helioseismology data
we use measurements of the solar surface radius from the ground and
space.

\section{Computations and updated results concerning the inversion of $f$-mode frequencies}

As in LK05, data used in this study are frequencies of solar
oscillation modes from 72 day MDI observing runs\footnote{These
files have been computed by J. Schou and are available on
\url{http://quake.stanford.edu/$\sim$schou/anavw72z/}}. Only
$f$-modes for the period 1996-2005 have been selected. The
oscillation modes, common for the whole period, are extracted to
finally obtain 148 modes ranging from $l=125$ to $l=285$. We will
here simply make a summary of the formalism used in LK05, and the
reader is invited to see that paper for detailed explanations.

A relation between the relative frequency variations $\delta\nu/\nu$
for $f$-modes and the associated Lagrangian perturbation of the
radius $\delta r/r$ of subsurface layers has been established by
\citet{Dziem04}:
\begin{equation}
\left(\frac{\delta\nu}{\nu}\right)_l=-\frac{3l}{2\omega^2 I}\int
dI\frac{g}{r}\frac{\delta r}{r} \label{eq_radius}
\end{equation}
where $l$ is the degree of the $f$-modes, $I$ is the moment of
inertia, $\omega$ is the eigenfrequency and $g$ is the gravity
acceleration. This equation is the starting point for our inversion
that will permit us to obtain the variation of the position of the
subsurface layers determined by $\delta r$ and $\delta\nu/\nu$. For
these calculations, we used  model S \citep{Christensen96}
calibrated to the seismic radius $R_{\odot}=6.9599\times10^5$ km,
and the standard Regularized Least-Square technique
\citep{Tikhonov77} (as we have an ill-posed inverse problem). For
the inversions, as in LK05, we select f-modes of angular degree
$\ell$ below 250, for which near-surface magnetic and turbulence
effects are not important \citep{Lefebvre06}.

The inversion of Eq. \ref{eq_radius} leads to the results presented
in Figure \ref{f1} which is an update of Figure 3 of LK05. The
figure presents the variation $\delta r$ for each year: every
difference has been computed relative to the reference year of 1996.
This updated graph presents the same features than the original one
published in LK05: 1) No significant changes in the variation of the
subsurface
    layers' depth below 0.97~$R_{\odot}$ with a maximal amplitude of
    about 10 km at $\simeq 0.985$~$R_{\odot}$; 2) Non-monotonic changes in
    the stratification, with the inner layer varying in phase
    with the solar activity cycle and
    the outer layer
    evolving almost in antiphase with the solar cycle with a maximal
    amplitude of about $26\pm 6$ km at about 0.995~$R_{\odot}$.

The new result is the radius changes are not precisely in antiphase
with the 11-year cycle. The results for 2005 indicate there might be
an additional phase lag between the radius change and activity. The
2005 data show the same behavior as 1997 data of the previous solar
minimum but the new activity minimum is not reached yet.
Significance of this result needs to be checked with future data; if
confirmed it may have interesting implications.

\section{Comparison of model predictions with the data}

In this section, we compare our results with the model results
obtained by SBDLT05. Figure 3 of SBDLT05 shows the radius change as
a function of depth below the photosphere relative to the radius
change at the depth of 5 Mm for their model 4, which includes both
temporal variation of magnetic field and turbulence. However, the
model does not provide an absolute value for the radius variations.
Therefore, we have to calibrate the model to obtain the best match
to the helioseismic f-mode measurements.


The principle is to use Eq. \ref{eq_radius} in a direct way by
assuming a value of $\left|\Delta R_{5Mm}\right|$, obtaining
$\frac{\delta r}{r}$ from model 4, and then calculating
$\frac{\delta\nu}{\nu}$ from Eq.~\ref{eq_radius}. Then, we search
for the $\left|\Delta R_{5Mm}\right|$ value, which provides the best
match to the observed changes of $f$-mode frequencies between 1996
and 2000. For the reference we use model S \citep{Christensen96}
calibrated with the seismic radius ($R_{\odot}=6.9599\times10^5$
km), the same as in the YREC model.  The first step was to digitize
the curve at the top of Figure 3 of SBDLT05 and transform to the
fractional radius. The panels of Figure \ref{f2} show: a) the
digitized curve as given by SBDLT05; b) the $\Delta R$ computed with
different $\left|\Delta R_{5Mm}\right|$ ranging from 0.3 to 1 km; c)
the associated $\Delta R/R$; and finally, d) computed
$\Delta\nu/\nu$ for the different $\Delta R/R$ (color curves) in
comparison with the $\Delta\nu/\nu$ observed between 1996 and 2000
(points with error bars).


Only few sample curves are plotted in panel d) of Figure \ref{f2},
and a study using the $\chi^2$ parameter shows that the minimum of
$\chi^2$ is obtained for $\Delta R_{5Mm}=0.65$ km, which gives a
value for $\Delta R$ at the surface of approximately 600 km.
However, such high variation of the solar radius has never been
observed in direct solar limb observations. In particular,
\citet{Kuhn04} put an upper limit of 7 mas, or just 5 km. This means
that the theoretical model of the solar-cycle variations cannot
satisfy both observational constraints, from helioseismology and
from the limb measurements. To fit the f-mode helioseismology data
it must have the surface radius variations by a factor 100 greater
than the observational upper limit from SOHO/MDI \citep{Kuhn04}. In
this case, it also exceeds the observational limit from ground-based
measurements by a factor of 4--20. Vice versa, if the model radius
is within the observational limit then the model cannot explain the
f-mode data.


Thus, we conclude that, according to the mathematical formalism
previously described, the strong non-homologous variation of the
subsurface solar radius proposed by SBDLT05 is not compatible with
the variations of the $f$-mode frequencies and the surface radius,
observed between 1996 and 2000. These computations explain why our
results published in LK05 and updated here are in contradiction in
amplitude and behavior with the results of SBDLT05. Of course our
computations are completely based on Eq. \ref{eq_radius} established
by \citet{Dziem04} and on the hypothesis that the temporal variation
of $f$-mode frequencies are mainly due to changes in the subsurface
stratification during the solar activity cycle. On the other hand,
the model proposed by SBDLT05 is based on the YREC code
\citep{Winnick02} into which the effects of magnetic fields and
turbulence have been included. But, as they said, the location,
magnitude, and temporal behavior of the internal field are not known
which imply to make assumptions on their treatment. We suggest that
the difference obtained with our results could come from their
treatment of the turbulence and magnetic fields in the subsurface
layers. Overcoming this discrepancy between the theory and
observations will help to understanding of the complicated magnetic
and turbulence effects below the visible surface of the Sun.

\section{Discussion}

\begin{figure}
\includegraphics[width=16cm]{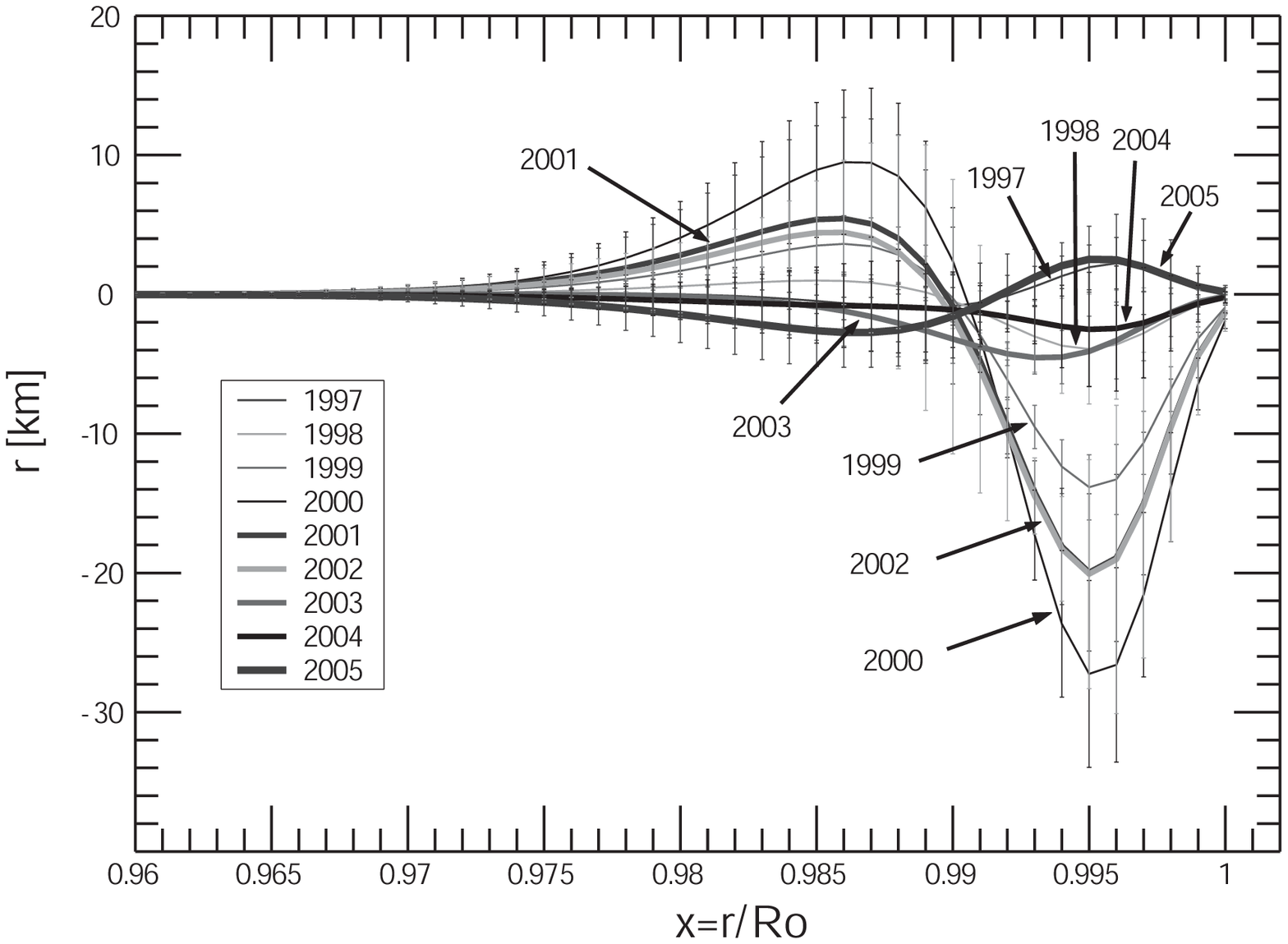}
\caption{ Radial variation $\left\langle \delta r
\right\rangle$ as a function of the fractional radius
$x=r/R_{\odot}$, obtained as a solution of the inversion of $f$-mode
frequencies by a least-squares regularization technique. The
reference year is 1996. The error bars are the standard deviation
after average over a set of random noise added to the relative
frequencies. The averaging kernels for this inversion are well
localized between 0.985 and 0.996, with a typical half-width of
about 0.003.} \label{f1}
\end{figure}

\begin{figure}
\includegraphics[width=16cm]{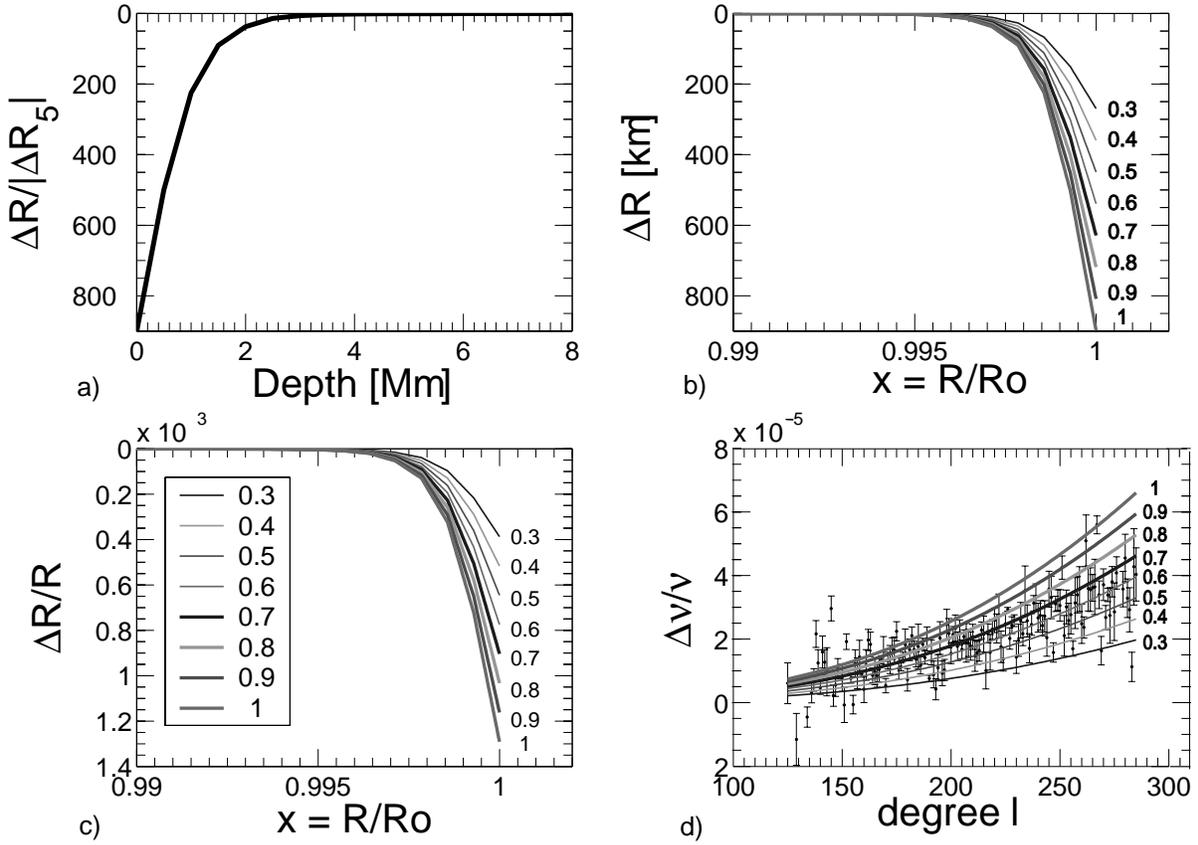}
\caption{\textit{a)}: The ratio between radius change as a
function of depth below the photosphere and radius change at 5 Mm as
a function of depth in Mm, i.e. same figure as in top panel of
Figure 3 of \citet{Sofia05}. \textit{b)}: Computed $\Delta R$ with
different $\left|\Delta R_{5Mm}\right|$ (see the legend expressed in
km in panel \textit{c}) as a function of the fractional radius
$x=R/R_{\odot}$. \textit{c)}: Computed $\Delta R/R$ with different
$\Delta R_{5Mm}$ from legend (in km) as a function of the fractional
radius $x$. \textit{d)}: Observed $\Delta\nu/\nu$ in black point
with errorbars and computed integrated $\Delta\nu/\nu$ as a function
of the degree $l$, using Eq. \ref{eq_radius} and $\Delta R/R$ from
left bottom panel.} \label{f2}
\end{figure}

In this paper, we have shown that the solar-cycle variations of the
solar radius that increase by a factor of approximately 1000 from a
depth of 5 Mm to the solar surface  as published by \citet{Sofia05}
(SBDLT05) are not consistent with the $f$-mode frequencies
variations between 1996 and 2000 and the upper limit on the solar
radius change from the simultaneous limb measurements. Using their
model and the relationship between $f$-mode frequencies and solar
radius variations, we obtained a variation at the solar surface of
about 600 km which is observed neither with ground-based nor space
instruments. The model of \citet{Sofia05} includes the effects of a
variable dynamo magnetic field and of a field-modulated turbulence
with general assumptions on their magnitude and temporal behavior
(based on an observed luminosity change of 0.1\% and the shape of
the activity cycle) because of the unknown location, magnitude and
temporal behavior of these internal fields. Nevertheless, we are
still cautious on our above results because (i) the near-surface
effects of turbulence and magnetic fields are not treated in our
approach and (ii) the lack of very-high $f$-mode degrees are
necessary to understand the very outer layers (approximately above 3
Mm). So we do not exclude a bigger variation of the position of
these layers than those plotted in Figure \ref{f1} and that the
solar radius at the photosphere by limb observations (as measured by
present instruments and future astrometric satellites) could be
larger than the seismic radius issued from $f$-mode frequencies
studies. Hence, we agree with SBDLT05 and the necessity to measure
precisely the variations of the solar radius from space.

\acknowledgments

This work utilizes data from SOHO/MDI and we thank J. Schou for
providing the frequencies files. SOHO is a project of international
cooperation between ESA and NASA. MDI is supported by NASA grant NAG
5-10483 to Stanford University. S. Lefebvre is supported by a CNES
post-doctoral grant in SAp. She would like to thank Sylvaine
Turck-Chi\`eze for useful discussions on this field. The authors
wish also to cordially thank both referees for very useful
suggestions on the manuscript.

\clearpage

\clearpage

\end{document}